\documentclass[a4paper,oneside,11pt]{article}
\usepackage{cprform,macros}
\usepackage{amsmath,amstext,amsfonts,amsbsy,amssymb,amscd,bbm,epsfig,lscape}

\newcommand{\ba}{\begin{array}}
\newcommand{\ea}{\end{array}}

\newcommand{\req}[1]{Eq.~(\ref{#1})}

\newcommand{\refig}[1]{Fig.~\ref{#1}}
\newcommand{\ret}[1]{Table~\ref{#1}}

\newcommand{\dif}{{\rm d}}

\newcommand{\Id}{\mathbf{1}}

\newcommand{\ci}[1]{\boldsymbol{#1}}

\newcommand{\Dslash}{\relax{\kern+.25em / \kern-.70em D}}

\newcommand{\Real}{\relax{\mathsf{\Gamma\kern-.35em R}}}
\newcommand{\Int}{\relax{\mathsf{Z\kern-.40em Z}}}

\newcommand{\half}{{\scriptstyle{{1\over 2}}}}

\newcommand{\MSbar}{{\overline{\rm MS}}}

\newcommand{\gbar}{\kern1pt\overline{\kern-1pt g\kern-0pt}\kern1pt}
\newcommand{\gren}{g_{\rm R}}

\newcommand{\mbar}{\kern2pt\overline{\kern-1pt m\kern-1pt}\kern1pt}

\newcommand{\mren}[1]{m_{{\rm R} #1}}
\newcommand{\obar}[1]{\kern3pt\overline{\kern-2pt #1\kern-0pt}\kern1pt}

\newcommand{\corrbar}[1]{\kern3pt\overline{\kern-2pt #1\kern-0pt}\kern1pt}

\newcommand{\orgi}[1]{\hat{#1}}

\newcommand{\hopc}{\kappa_{\rm cr}}

\newcommand{\oVApAVren}[1]{\kern3pt\overline{\kern-2pt #1\kern-0pt}\kern1pt_{\rm\scriptscriptstyle VA+AV;s}}

\newcommand{\ZA}{Z_{\rm\scriptscriptstyle A}}
\newcommand{\ZV}{Z_{\rm\scriptscriptstyle V}}

\newcommand{\zbar}{\kern3pt\overline{\kern-2pt Z\kern-0pt}\kern1pt}
\newcommand{\zrgi}{\hat Z}
\newcommand{\zbarVApAV}[1]{\kern3pt\overline{\kern-2pt Z\kern-0pt}\kern1pt_{\rm\scriptscriptstyle VA+AV #1}}

\newcommand{\Oa}{\mbox{O}(a)}
\newcommand{\Oasq}{\mbox{O}(a^2)}

\newcommand{\icA}{c_{\rm\scriptscriptstyle A}}

\newcommand{\cH}{{\cal H}}

\newcommand{\cQ}{{\cal Q}}
\newcommand{\cR}{{\cal R}}

\newcommand{\eq}[1]{Eq.~(\ref{#1})}
\long\def\symbolfootnote[#1]#2{\begingroup%
\def\thefootnote{\fnsymbol{footnote}}\footnote[#1]{#2}\endgroup}

\begin{document}


\begin{titlepage}


\vspace*{-30truemm}
\begin{flushright}
ROM2F/2006-17,
CERN-PH-TH/2006-131,
FTUV-06-2007\\
IFIC/06-29,
MKPH-T-06-14,
DESY 06-112\\
July 2006
\end{flushright}
\vspace{15truemm}


\centerline{\Bigrm Non-perturbative renormalisation of left-left}
\vskip 2 true mm
\centerline{\Bigrm four-fermion operators with Neuberger fermions}
\vskip 9 true mm
\vskip -2 true mm
\centerline{\bigrm P.~Dimopoulos$^a$\symbolfootnote[2]{Member of the ALPHA Collaboration}, L.~Giusti$^b$\symbolfootnote[1]{On leave from Centre de Physique Th\'eorique, CNRS Luminy, F-13288 Marseille, France.}, P.~Hern\'andez$^c$, F.~Palombi$^{d\dagger}$,}
\centerline{\bigrm C.~Pena$^{b\dagger}$, A.~Vladikas$^{a\dagger}$, J.~Wennekers$^e$ and H.~Wittig$^{d\dagger}$}
\vskip 4 true mm
\centerline{\it $^a$ INFN, Sezione di Roma ``Tor Vergata''}
\centerline{\it c/o Dipartimento di Fisica, Universit\`a di Roma ``Tor
  Vergata''}
\centerline{\it Via della Ricerca Scientifica 1, I-00133 Rome, Italy}
\vskip 3 true mm
\centerline{\it $^b$ CERN, Physics Department, TH Division}
\centerline{\it CH-1211 Geneva 23, Switzerland}
\vskip 3 true mm
\centerline{\it $^c$ Dpto. de F\'{\i}sica Te\'orica and IFIC, Universitat de Val\`encia}
\centerline{\it E-46100 Burjassot, Spain}
\vskip 3 true mm
\centerline{\it $^d$ Institut f\"ur Kernphysik, University of Mainz}
\centerline{\it D-55099 Mainz, Germany}
\vskip 3 true mm
\centerline{\it $^e$ DESY, Theory Group}
\centerline{\it Notkestra\ss e 85, D-22603 Hamburg, Germany}
\vskip 10 true mm


\thintablerule
\vskip 3 true mm
\noindent{\tenbf Abstract}
\vskip 1 true mm
\noindent
{\tenrm
We outline a general strategy for the
non-perturbative renormalisation of composite operators in discretisations based on Neuberger fermions, via a matching to results obtained 
with Wilson-type fermions. As an application, we consider the 
renormalisation of 
the four-quark  operators entering the $\Delta S=1$ and $\Delta S=2$ 
effective Hamiltonians. 
Our results are an essential ingredient for the determination of the low-energy constants governing non-leptonic kaon 
decays.
}
\vskip 3 true mm
\thintablerule
\vspace{10truemm}

\end{titlepage}
\eject

\section{Introduction}

The renormalisation of four-fermion operators is an essential ingredient in lattice QCD computations of weak matrix
elements. In this letter we will address the logarithmically divergent
renormalisation of left-left four-quark operators, with an
emphasis on the $\Delta S=1$ effective Hamiltonian
governing non-leptonic kaon decays.

The treatment of $\Delta S=1$ weak decays via an
effective weak Hamiltonian with an active charm
quark has been recently reviewed in~\cite{Giusti:2004an}. After
performing the operator product expansion and
neglecting top quark effects, which are suppressed
by three orders of magnitude relative to the contributions of up and charm quarks, the expression found for the $\Delta S=1$
effective weak Hamiltonian in the formal continuum
theory is
\begin{gather}
\label{eq:Hw}
\cH_{\rm w} = \frac{g_{\rm w}^2}{4M_W^2}(V_{us})^*V_{ud}
\sum_{\sigma=\pm}\{k_1^\sigma\cQ_1^\sigma + k_2^\sigma\cQ_2^\sigma\} \, .
\end{gather}
In the above expression
$g_{\rm w}=4\sqrt{2}G_{\rm F}M_W^2$, $k_{1,2}^\sigma$ 
are Wilson coefficients, and the dimension-six operators
$\cQ_{1,2}^\sigma$ have the form
\begin{align}
\cQ_1^\pm &=
[(\bar s\gamma_\mu P_- u)(\bar u\gamma_\mu P_- d) \pm
 (\bar s\gamma_\mu P_- d)(\bar u\gamma_\mu P_- u)]
 -[u~\rightarrow~c] \, ,\\
\cQ_2^\pm &=
(m_u^2-m_c^2)\{m_d(\bar s P_+ d)+m_s(\bar s P_- d)\} \, ,
\end{align}
where parentheses around quark bilinears indicate colour
and spin traces and $P_\pm =\half(\Id\pm\gamma_5)$. Although our procedure is completely general, we will from
now on concentrate in the
${\rm SU}(4)_{\rm L} \times {\rm SU}(4)_{\rm R}$ symmetric
limit, where all quark masses are degenerate~\cite{Giusti:2004an}. In this limit
the only contribution to decay amplitudes comes from matrix
elements of the operators $\cQ_1^\pm$. Moreover, 
the operator renormalisation pattern is greatly
simplified, as mixing with lower dimension operators is absent.
We stress, however, that our results, being obtained in a
mass-independent renormalisation scheme, will renormalise
properly subtracted operators also beyond the 
${\rm SU}(4)_{\rm L} \times {\rm SU}(4)_{\rm R}$ symmetric
limit.

Our strategy to renormalise $\cQ^\pm_1$ is similar
to the technique proposed in~\cite{Hernandez:2001yn} for the computation of
the renormalised chiral condensate. It involves matching
bare correlation functions (or matrix elements) computed with 
Neuberger fermions to their renormalisation group invariant (RGI)
counterparts. The latter are computed in the continuum limit with some variant
of Wilson fermions, for which mature techniques for fully
non-perturbative renormalisation exist. Our choice will be twisted 
mass QCD (tmQCD) with an $\Oa$ improved fermion action.

Although we will concentrate specifically
on the operators of the $\Delta S=1$ Hamiltonian, both the proposed
methodology and our results have a wider range of application.
In particular, the renormalisation factors that we will obtain
for $\cQ^+_1$ renormalise also the four-fermion operator entering
the $\Delta S=2$ effective Hamiltonian.
In the present work, all computations are performed in the 
quenched approximation.

In the next section we will describe the strategy of the computation.
In section~3 we discuss the computation of RGI operators in the
continuum limit, based on a twisted mass QCD (tmQCD) Wilson fermion regularisation. In section~4 we
discuss our results with Neuberger fermions, and compute non-perturbative 
renormalisation factors. Section~5 deals with
perturbative estimates of the same renormalisation factors. We present
our conclusions in section~6.

\section{Strategy}

Let us consider a generic multiplicatively renormalisable operator.\footnote{The generalisation to operators which mix under renormalisation is straightforward.}
The notation will follow closely that of~\cite{Giusti:2004an}. We will be dealing only with mass-independent
renormalisation schemes. 

We start by recalling the definition of renormalisation group
invariant (RGI) composite operators. The RGI insertion of a local operator 
$\cQ$
into a continuum on-shell correlation function is given by
\begin{gather}
\begin{split}
\orgi{\cQ}(\gren,\mren{},L) &= c_{\rm s}(\mu/\Lambda)\, \lim_{a \rightarrow 0}
Z_{\rm s}(g_0,a\mu)\,\cQ(g_0,m_0,L) \\
&\equiv \lim_{a \rightarrow 0} \zrgi(g_0)\,\cQ(g_0,m_0,L) \, ,
\label{eq:Qrgi}
\end{split}
\end{gather}
where $Z_{\rm s}$ is a renormalisation constant that renders the operator
finite, $\Lambda$ is the QCD scale, and $g_0,m_0$ ($\gren,\mren{}$) denote the bare
(renormalised) gauge coupling and quark mass(es). The subscript ``${\rm s}$'' labels the renormalisation scheme.
We have also indicated
explicitly that correlation functions will be computed in a finite volume of 
spatial size $L$ (eventually taking $L\to\infty$).
The RG-evolution function is given by
\begin{gather}
c_{\rm s}(\mu/\Lambda) =
\left[2b_0\gbar^2(\mu)\right]^{\gamma_0/(2b_0)}
\exp\left\{-\int_0^{\gbar(\mu)}\dif g\left[\frac{\gamma(g)}{\beta(g)}
+\frac{\gamma_0}{b_0 g}\right]\right\} \, ,
\end{gather}
where we have used the perturbative expansions of
the anomalous dimension of the operator $\gamma$ and the $\beta$-function, viz.
\begin{gather}
\beta(g) \stackrel{g \to 0}{\approx} -g^3(b_0+b_1 g^2+\ldots) \, ,~~~~~~~
\gamma(g) \stackrel{g \to 0}{\approx} g^2(\gamma_0+\gamma_1 g^2+\ldots) \, .
\end{gather}
It has to be stressed that
the (scale-independent) RGI renormalisation factor
$\zrgi(g_0) \equiv c_{\rm s}(\mu/\Lambda) Z_{\rm s}(g_0,a\mu)$
depends on the renormalisation scheme only via
cutoff effects, since the RGI operator insertion $\orgi{\cQ}$ is scheme-independent. On the other hand, $\zrgi$ is
regular\-isa\-tion-dependent. We also stress that the running factor
$c_{\rm s}$ is a continuum quantity, and hence regularisation-independent.

We now consider two different
lattice regularisations, namely Wilson (denoted by ``w'') and Neuberger (or overlap)
fermions (denoted by ``ov''). Our aim is to construct RGI renormalisation
factors for Neuberger fermions operators by matching renormalised quantities
in both regularisations.
The first step consists of using the
first regularisation in order to compute the RGI operator $\orgi{\cQ}$
at a reference physical point, parametrised here by
$({\gren}_{,\rm ref},\mren{,\rm ref},L_{\rm ref})$, viz.
\begin{gather}
\label{eq:RGIw_def}
\orgi{\cQ}({\gren}_{,\rm ref},\mren{,\rm ref},L_{\rm ref}) = \lim_{a \to 0}
\zrgi^{\rm w}(g_0)\,\cQ^{\rm w}(g_0,m_0,L_{\rm ref}) \, .
\end{gather}
It is essential to note that any reference to the regularisation
employed in the r.h.s. of eq.~(\ref{eq:RGIw_def}) has disappeared
after the continuum limit has been taken. The second step consists of tuning a point
$(g'_0,m'_0)$ in the bare parameter space of the second regularisation,
which corresponds to the same values of the renormalised parameters 
$({\gren}_{,\rm ref},\mren{,\rm ref})$. Assuming universality of the continuum limit, one
then has
\begin{gather}
\begin{split}
\label{eq:RGIov_def}
\orgi{\cQ}({\gren}_{,\rm ref},\mren{,\rm ref},L_{\rm ref}) &= \zrgi^{\rm ov}(g_0')\,\cQ^{\rm ov}(g'_0,m'_0,L_{\rm ref}) \, + \Oasq \, ,
\end{split}
\end{gather}
where we have explicitly used the fact that correlation functions
computed with Neuberger fermions exhibit scaling violations of at most $\Oasq$.
Once the bare quantity $\cQ^{\rm ov}(g'_0,m'_0,L_{\rm ref})$ has
been computed, eq.~(\ref{eq:RGIov_def}) yields $\zrgi^{\rm ov}(g_0')$, provided that $\orgi{\cQ}({\gren}_{,\rm ref},\mren{,\rm ref},L_{\rm ref})$ has been determined through eq.~(\ref{eq:RGIw_def}).
This procedure
can be repeated at several bare couplings and masses $(g'_0,m'_0)$,
always corresponding to $({\gren}_{,\rm ref},\mren{,\rm ref})$;
in this way the lattice spacing may be varied, while the physics (i.e. physical volume and
renormalised coupling and masses) is kept fixed.
Note that \eq{eq:RGIov_def} is to be
interpreted as a renormalisation condition that implicitly defines
a mass-independent renormalisation scheme. Thus it ensures that the RGI renormalisation factors computed in this way will correctly
renormalise the operators at any value of the quark masses $\mren{}$. 
In particular, $\zrgi^{\rm ov}$ depends on quark masses only via cutoff effects
(though it is not guaranteed a priori that such dependence is small).
On the other hand, the renormalisation prescription~(\ref{eq:RGIov_def}) reproduces by construction the RGI result at the reference point for any chosen set of bare parameters. Thus the procedure is only useful if the targeted physical regime, characterised by $(\gren,\mren{},L)$, is well away from $({\gren}_{,\rm ref},\mren{,\rm ref},L_{\rm ref})$.

The present work provides an application of this strategy.
The ultimate aim, which is achieved in ref.~\cite{prl}, is the computation of the
effective low-energy couplings governing non-leptonic kaon decays,
following the strategy described in~\cite{Giusti:2004an}. This involves the computation, carried out using Neuberger fermions, of the chiral limit values of the ratios of correlation functions
\begin{gather}
\label{eq:ratioLL}
R^{\pm}_1(x_0,y_0)=
\frac{\langle [J_0(x)]_{du}\cQ_1^\pm(0)[J_0(y)]_{us}\rangle}
{\langle [J_0(x)]_{du} [J_0(0)]_{ud} \rangle
 \langle [J_0(0)]_{su} [J_0(y)]_{us} \rangle} \, ,
\end{gather}
where $J_\mu$ is the left-handed current
\begin{gather}
[J_\mu(x)]_{\alpha\beta} = \bar \psi_\alpha \gamma_\mu P_- \psi_\beta \, ,
\end{gather}
and $\alpha,\beta$ are flavour indices. An essential ingredient of the procedure are the renormalisation factors $[\zrgi_1^{\pm}/\ZA^2]$,
the computation of which is the goal of the present work.

In order to compute non-perturbatively the renormalisation factors $[\zrgi_1^{\pm}/\ZA^2]$ for Neuberger fermions, we will employ the ratios of QCD matrix elements
\begin{gather}
\label{eq:ratWME}
\cR_\pm \equiv
\frac{\langle \pi^+|\cQ_1^\pm| K^+ \rangle}
{\langle \pi^+|J_0|0\rangle\langle 0|J_0|K^+\rangle} \,
\end{gather}
computed in large volumes and at a value of the reference quark
mass $\mren{,{\rm ref}}$ corresponding to $m_{\rm PS}=m_\pi=m_K=m_K^{\rm phys} = 495~\MeV$.
Note that, in the ${\rm SU}(4)_{\rm L} \times {\rm SU}(4)_{\rm R}$
symmetric limit,
the ratio $\cR_+$ will be equal, up to a trivial factor, to the Kaon bag parameter
$B_K$. 
The RGI ratios $\orgi{\cR}_\pm$ will be computed,
as in \eq{eq:RGIw_def}, using a Wilson fermion regularisation.
To this purpose, we will compute the bare quantities $\cR_\pm$
at several values of the bare coupling $g_0$, and use the RGI 
renormalisation factors computed in the same $g_0$ range in~\cite{Guagnelli:2005zc}, using a 
Schr\"odinger Functional (SF) framework.
We will then apply \eq{eq:RGIov_def} to match the RGI ratios 
$\orgi{\cR}_1^\pm$ to the ratios of bare matrix elements computed with Neuberger fermions.
Since the matching reference regime of large volumes and meson masses of the
order of $m_K^{\rm phys}$ is well away from the target one in 
which $[\zrgi^{\pm}/\ZA^2]$
are to be used, the construction is indeed
non-trivial. In this way we have exploited the fact that
Wilson fermions are well suited for simulations in the
strange-quark regime, while they become problematic close
to the chiral limit, where Neuberger fermions are clearly
advantageous.

For the sake of consistency, we will also perform
a direct determination of the ratio
$\zrgi_1^{+;\rm ov}/\zrgi_1^{-;\rm ov}$, computed from
a matching involving the ratio of matrix elements
\begin{gather}
\label{eq:ratWME2}
\frac{\cR_+}{\cR_-} =
\frac{\langle \pi^+|\cQ_1^+| K^+ \rangle}
     {\langle \pi^+|\cQ_1^-| K^+ \rangle} \, .
\end{gather}
This specific ratio is of particular interest, as it enters
directly the study of the $\Delta I=1/2$ enhancement rule.

Some comments are in order. The renormalisation factors obtained
via the procedure just described do not lead, obviously,
to independent renormalised values for the ratios of matrix elements
in \eq{eq:ratWME} computed at the reference point $m_K^{\rm phys}$, as a tautology would result. As explained
above, the renormalisation factors $[\zrgi_1^{\pm}/\ZA^2]$ will 
rather be used to renormalise quantities effectively computed in the chiral limit.
They can also be used e.g. to renormalise ratios of correlation
functions computed in the $\epsilon$-regime. The fact that the matching involves correlation functions computed with both 
periodic and SF boundary conditions does not, on the other hand, 
give rise to any subtlety, as only hadronic matrix elements
computed in a large volume are involved.

Unlike the above strategy, adopted in this work, the ideal matching
procedure should not involve the tradeoff of a long-distance
matrix element of physical relevance.
It is e.g. possible to match a different matrix element
of the same operator, or to take a reference point for the
matching which is well away from all the target physical regimes of
interest. On the other hand, the particular strategy adopted
here has the advantage that it allows to use the numerical
results obtained in the context of~\cite{Dimopoulos:2006dm}.

\section{Wilson-tmQCD computation of RGI operators}

We will now discuss the computation of the RGI
operators $\orgi{\cQ}_1^\pm$ (i.e. the l.h.s. of \req{eq:RGIov_def}),
using the tmQCD formalism with Wilson quarks~\cite{Frezzotti:2000nk}.

We start by recalling that, with Wilson fermions, the renormalisation of $\cQ_1^\pm$ is more subtle than in chirally symmetric regularisations (see \cite{Donini:1999sf} for a detailed discussion). Customarily, both operators are split into
parity-even and parity-odd parts as
\begin{gather}
\cQ_1^\pm = \cQ^\pm_{\rm VV+AA} - \cQ^\pm_{\rm VA+AV} \, ,
\end{gather}
in standard notation. In the three-point correlation functions considered below, parity conservation in QCD ensures that
the only contribution comes from the parity-even part
$\cQ^\pm_{\rm VV+AA}$. With ordinary Wilson
fermions, as a consequence of the breaking of chiral symmetry, the
renormalisation of $\cQ^\pm_{\rm VV+AA}$ requires the subtraction
of four finite counterterms involving all the remaining
Lorentz-invariant, parity-even four-quark operators with the same
flavour structure.\footnote{Mixing with operators of lower dimension
is always absent in the ${\rm SU}(4)_{\rm L} \times {\rm SU}(4)_{\rm R}$ limit, as all mixing coefficients are proportional to the mass difference $(m_c-m_u)$.} On the other hand, it is possible to 
construct
twisted mass QCD (tmQCD) Wilson regularisations in which the
counterterms are absent, and the operator renormalises
multiplicatively. The basic property that has to be
satisfied is that the chiral rotation of the quark fields
that generates the twisted mass term maps $\cQ^\pm_{\rm VV+AA}$ onto
$\cQ^\pm_{\rm VA+AV}$. In mass-independent renormalisation schemes,
the latter is protected from mixing with other four-fermion operators 
by CPS symmetry.
Examples of such regularisations have been discussed in~\cite{Pena:2004gb}.

Here we will adopt, however, a different approach,
which will allow us to use the numerical results obtained in~\cite{Dimopoulos:2006dm}. To that purpose we restrict ourselves
to the quenched approximation, and use the formalism for valence
quark flavours advocated in~\cite{Frezzotti:2004wz}. We consider a theory
with six valence flavours, that we will label 
$\psi=(u,d,s,c,u',c')^T$. 
We will employ two tmQCD regularisations, characterised by the choice of twist angle $\alpha$ in the definition of the fermion action:
\begin{gather}
S^{(\alpha)}_{\rm tmQCD} = a^4 \sum_{x,y}\bar\psi(x)\left\{D_{\rm w,sw}
+ \mathbf{m}^{(\alpha)} + i\gamma_5\ci{\mu}^{(\alpha)}\right\}(x,y)\,\psi(y) \, ,
\end{gather}
where $D_{\rm w,sw}$ is the Wilson operator with a
Sheikholeslami-Wohlert term, $\mathbf{m}^{(\alpha)}$ and $\ci{\mu}^{(\alpha)}$ are diagonal mass matrices, and the label $\alpha$ refers
to the twist angle entering chiral rotations.
The bare mass
parameters are tuned so that, up to $\Oasq$ corrections, the
renormalised mass matrices have the form
\begin{alignat}{3}
\mathbf{m}^{(\pi/2)}_{\rm R} &= M_{\rm R}\,{\rm diag}(0,0,1,0,1,1) \, ,~~~~
&\ci{\mu}^{(\pi/2)}_{\rm R} &= M_{\rm R}\,{\rm diag}(1,1,0,1,0,0) \, , \\
\mathbf{m}^{(\pi/4)}_{\rm R} &= \frac{M_{\rm R}}{\sqrt{2}}\,{\rm diag}(1,1,1,1,1,1) \, ,~~~~
&\ci{\mu}^{(\pi/4)}_{\rm R} &= \frac{M_{\rm R}}{\sqrt{2}}\,{\rm diag}(1,1,-1,1,-1,-1) \, ,
\end{alignat}
where $M_{\rm R}$ is the physical renormalised quark mass.
The mass tuning procedure
is identical to the one described in~\cite{Dimopoulos:2006dm}.
We then introduce the operators
\begin{gather}
\begin{split}
\widetilde\cQ_1^\pm =
&[(\bar s\gamma_\mu P_- u)(\bar u'\gamma_\mu P_- d) \pm
 (\bar s\gamma_\mu P_- d)(\bar u'\gamma_\mu P_- u)]
 -[u\rightarrow c\,,\,u'\rightarrow c'] \, .
\end{split}
\end{gather}
Using the standard relation between QCD and tmQCD in the continuum 
limit, one finds, for the two regularisations specified above, that 
the following equalities hold between operator insertions in RGI 
correlation functions 
\begin{gather}
[\orgi{\widetilde\cQ}_{\rm VV+AA}^\pm]_{\rm QCD} =
i \,[\orgi{\widetilde\cQ}_{\rm VA+AV}^\pm]_{\rm tmQCD} \, ,\\
\ZA[A_\mu]_{su,{\rm QCD}} =
\frac{1}{\sqrt{2}}\left\{
\ZA[A_\mu]_{su,{\rm tmQCD}}-i\ZV[V_\mu]_{su,{\rm tmQCD}}
\right\} \, ,\\
\ZA[A_\mu]_{du',{\rm QCD}} =
\frac{1}{\sqrt{2}}\left\{
\ZA[A_\mu]_{du',{\rm tmQCD}}-i\ZV[V_\mu]_{du',{\rm tmQCD}}
\right\} \, ,
\end{gather}
where $[A_\mu]_{\alpha\beta}=\bar\psi_\alpha\gamma_\mu\gamma_5\psi_\beta$
and $[V_\mu]_{\alpha\beta}=\bar\psi_\alpha\gamma_\mu\psi_\beta$.
The current normalisations $\ZA,\ZV$, as well as the $\Oa$
improvement coefficient $\icA$ needed to construct an $\Oa$ improved axial current, are set to the values computed by
the Alpha Collaboration~\cite{imprAlpha1,imprAlpha2}.

Correlation functions are computed within a Schr\"odinger functional
(SF) framework, with quark and gluon fields obeying periodic boundary
conditions in space (with period $L$) and (homogeneous) Dirichlet
boundary conditions in time at the hypersurfaces $x_0 = 0$ and $x_0 =
T$. Ratios of correlation functions, from which the ratios of matrix
elements $\cR_\pm$ and $\cR_+/\cR_-$ can be extracted, are defined in complete analogy to the ones specified for the extraction of $B_K$ in~\cite{Dimopoulos:2006dm}.
Our run parameters, too, are the same as in~\cite{Dimopoulos:2006dm}, 
and are listed in Tables~1 and~5 of that work, save for one important exception, concerning the dataset at $\beta=6.1$. After completion
of ref.~\cite{Dimopoulos:2006dm}, its authors carried out a proper
determination of $\hopc(\beta=6.1)$, based on the method of~\cite{imprAlpha1}. They found that this estimate of $\hopc$ disagrees by several standard deviations from the value cited
in the literature~\cite{Rolf:2002gu}, which had been used in~\cite{Dimopoulos:2006dm} for the tuning of the bare mass parameters. This discrepancy hence necessitated a new determination of the bare mass parameters in order to satisfy the constraints imposed by the prescribed values of the twist angles. Consequently, the runs at $\beta=6.1$ had to be repeated after completion of~\cite{Dimopoulos:2006dm}. Full details will be provided separately~\cite{BK_erratum}. In the present work we merely quote the corrected value of $\cR_+$ in the continuum limit.

\begin{figure}
\begin{center}
\vspace{19.0cm}
\includegraphics{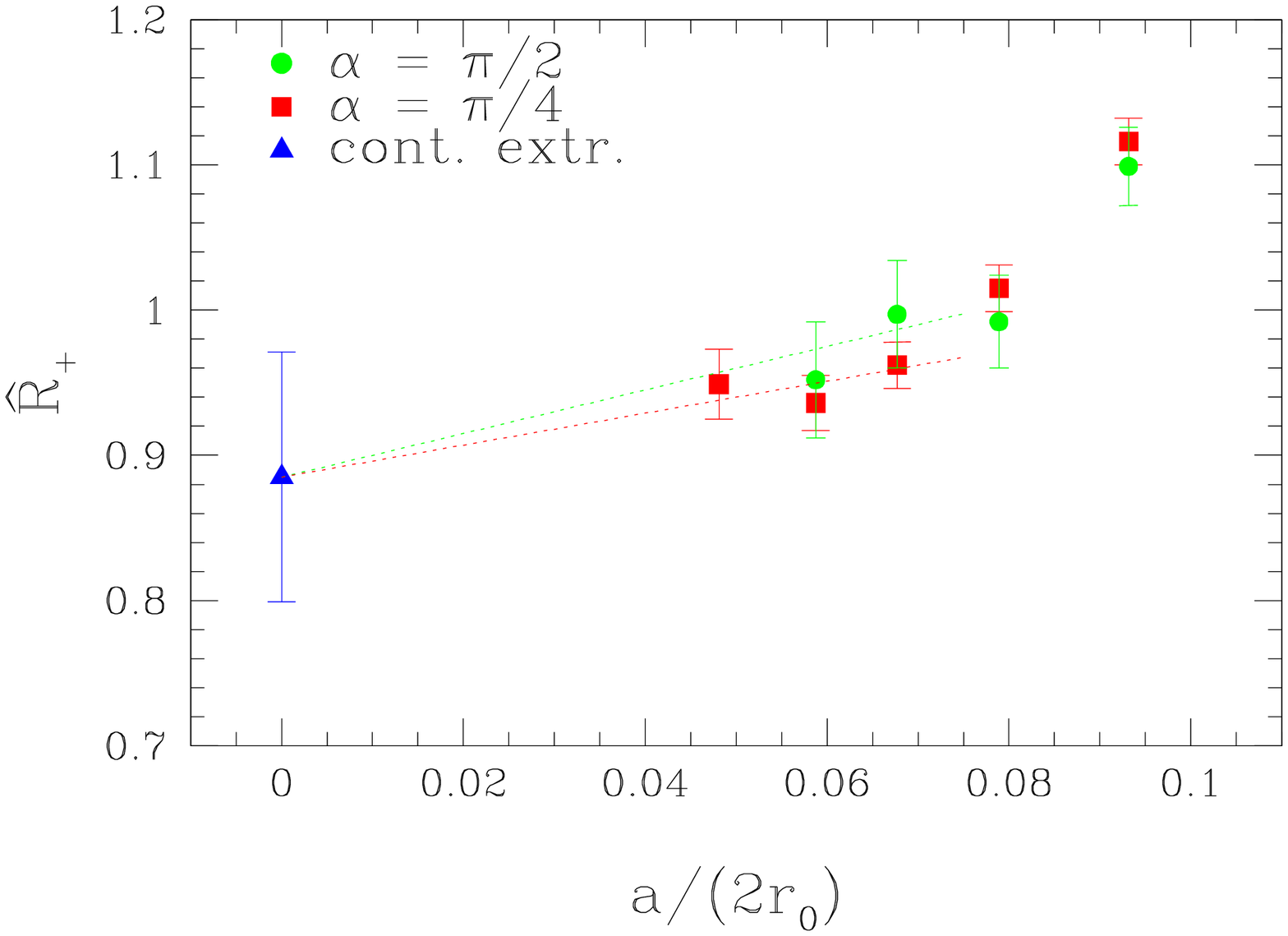}
\includegraphics{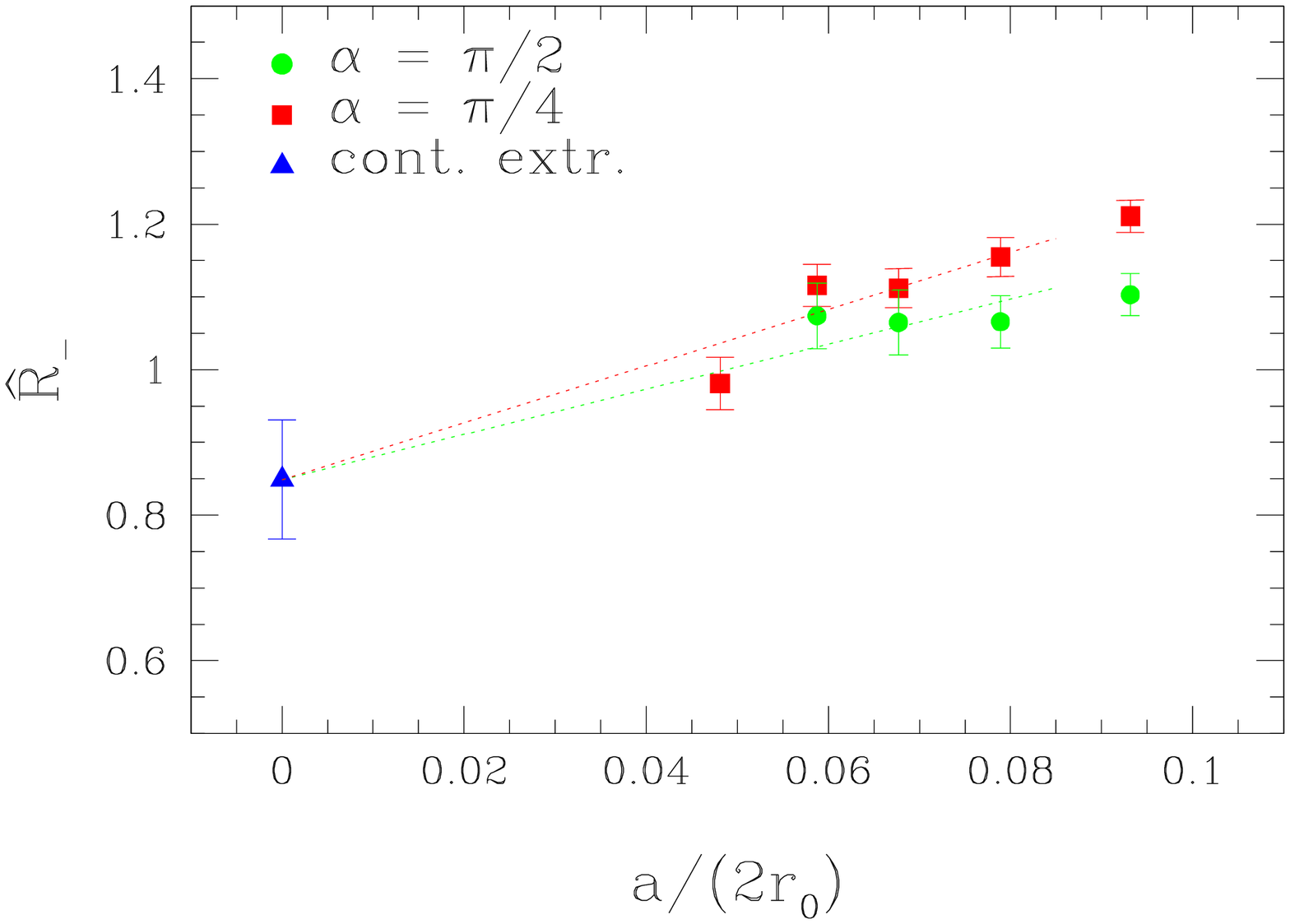}
\includegraphics{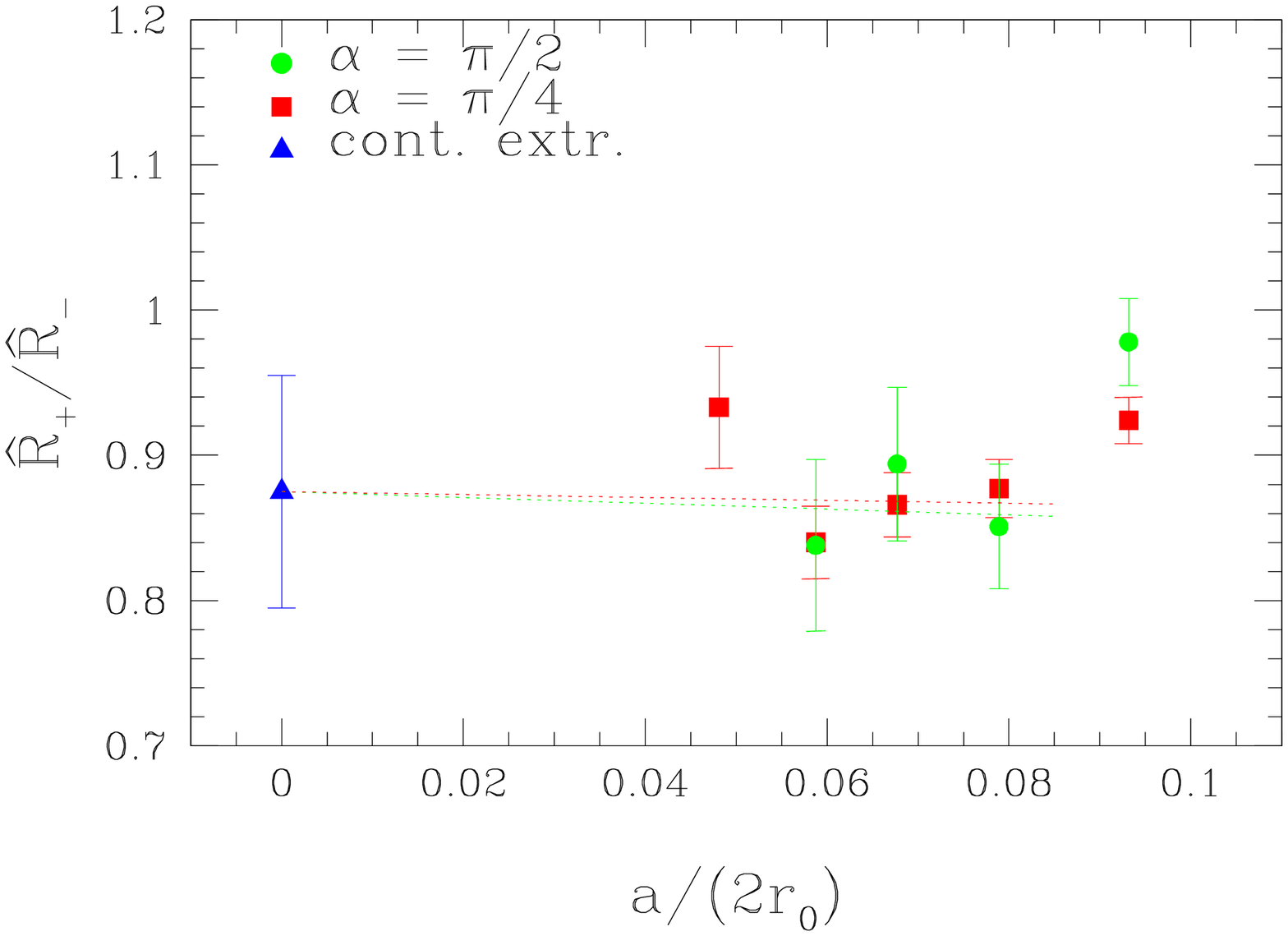}
\caption{Continuum limit extrapolations of $\orgi{\cR}^\pm$ and $\frac{\orgi{\cR}^+}{\orgi{\cR}^-}$.}
\label{fig:CLextrap}
\end{center}
\end{figure}

The RGI ratios are
obtained upon multiplication by the RGI renormalisation factors
$\zrgi_{\rm VA+AV}^\pm$. The latter have
been computed non-perturbatively in~\cite{Guagnelli:2005zc},
using the standard SF finite-size scaling analysis,
for a range of inverse couplings
$6.0 \lesssim \beta \lesssim 6.5$.
Out of the various SF renormalisation schemes considered
in~\cite{Guagnelli:2005zc} we have chosen to employ scheme 1 
for the renormalisation of $\widetilde\cQ_{\rm VA+AV}^+$ and scheme 8 for
$\widetilde\cQ_{\rm VA+AV}^-$; the reasons are explained in~\cite{Guagnelli:2005zc} and~\cite{Palombi:2005zd}.
The appropriate error analysis has been extensively discussed
(for $\orgi{\cR}_+$) in~\cite{Dimopoulos:2006dm}. The continuum limit
is then obtained by performing a combined extrapolation of 
the results coming from both tmQCD regularisations.
The extrapolation is linear in the
lattice spacing $a$, since the four-fermion operator is not $\Oa$ 
improved, and hence the leading lattice artifacts in $\cR_\pm$ 
are expected to be $\Oa$. Furthermore, as discussed in~\cite{Dimopoulos:2006dm}, at the lowest values of $\beta$
the $\Oa$ ambiguity in the determination of the improvement coefficient $\icA$ has a significant impact on cutoff effects.
Under these premises, our most stable continuum limit
extrapolation for $\orgi{\cR}_+$ is obtained by discarding
the $\beta=6.0,6.1$ data points, while for $\orgi{\cR}_-$
and $\orgi{\cR}_+/\orgi{\cR}_-$ only $\beta=6.0$ is discarded.
The final results, illustrated in \refig{fig:CLextrap}, are
\begin{gather}
\label{eq:RGI1}
\orgi{\cR}_+ = 0.885(86) \, ,~~~~~~~~~~~
\orgi{\cR}_- = 0.849(82) \, ,\\
\label{eq:RGI3}
\frac{\orgi{\cR}_+}{\orgi{\cR}_-} = 0.875(80) \, .
\end{gather}
We stress that the volume dependence of these results
is well within the quoted uncertainties (see~\cite{Dimopoulos:2006dm} for details).

Eqs.~(\ref{eq:RGI1}-\ref{eq:RGI3}) are the main result of the
present work. In the next section we will use them to determine the 
renormalisation factors needed with Neuberger fermions. It must be stressed at this point that the continuum limit extrapolation is rather long and, in the case of $\orgi{\cR}_-$, strongly driven by the $\beta=6.45$ datum. A better control of the continuum limit extrapolations could be achieved e.g. by removing $\Oa$ effects as suggested in ref~\cite{Frezzotti:2004wz}. This is beyond the scope of the present letter.

\section{Renormalisation constants for Neuberger fermions}

Having constructed the RGI ratios of matrix elements
in eq.~(\ref{eq:ratWME}), we now insert them in \req{eq:RGIov_def} and solve for the Neuberger fermions renormalisation
constants $\zrgi_1^{\pm,\rm ov}(g_0)$.

In order to regularise the theory using Neuberger fermions,
we start by introducing the Neuberger-Dirac operator~\cite{Neuberger:1997fp}
\begin{gather}
\label{eq:NeubergerDiracOp}
D = \frac{1}{\abar}\left\{\Id-A(A^\dagger A)^{-1/2}\right\}\, ,~~~~~~~
A = 1+s-aD_{\rm w} \, ,
\end{gather}
where $D_{\rm w}$ is the massless Wilson-Dirac operator, 
$a$ denotes the lattice spacing, and $s$ is a free parameter in the 
range $|s|<1$. By setting $\abar=a/(1+s)$ it is straightforward to 
check that $D$ satisfies the Ginsparg-Wilson relation
\begin{gather}
\gamma_5 D + D \gamma_5 = \abar D \gamma_5 D \, .
\end{gather}
Composite operators which have proper chiral
transformation properties in the regularised theory are obtained
by performing the substitution
\begin{gather}
\begin{align}
\psi & \to \left(\Id-\half\abar D\right)\psi \, ,
\nonumber \\
\bar \psi & \to \bar \psi \, .
\end{align}
\end{gather}
The operators $\cQ_1^\pm$
in the discretised theory share the same transformation properties
under chiral symmetry as their counterparts in the continuum
(see~\cite{Giusti:2004an} and references therein).

\begin{table}[!t]
\begin{center}
\begin{tabular}{cccccc}
\Hline \\[-10pt]
$\beta$ & $am$ & $r_0 m_{\rm PS}$& $\cR^{\rm ov}_+$ & $\cR^{\rm ov}_-$ & $\cR^{\rm ov}_+/\cR^{\rm ov}_-$ \\
&&&&& \\[-10pt]
\hline \\[-10pt]
$5.8485$ & $0.060$ & $1.259(10)$ & $0.772(30)$ & $1.514(73)$ & 0.511(28)
\\ \\[-10pt]
\Hline\\[-15pt]
\end{tabular}
\end{center}
\caption{Results with Neuberger fermions, obtained on a $16^3 \times 32$
lattice from 197 configurations and using low-mode
averaging with 20 low modes of the Neuberger-Dirac operator. The
physical spatial extent of the lattice is $L/(2r_0)=1.98$.}
\label{tab:overlap_results}
\end{table}

Bare values for the ratios of matrix elements $\cR_\pm$, are extracted 
from the ratios of correlation functions of eq.~(\ref{eq:ratioLL}).
The details of the computation, performed at a fixed value of $\beta$ 
with periodic boundary conditions
in all Euclidean spacetime directions, are reported
in~\cite{Giusti:2004an,Giusti:2005pd,prl}.
The simulation parameters and our results for $\cR_\pm$
are provided in \ret{tab:overlap_results}. Since our
pseudoscalar mass is compatible within errors with the
kaon mass $r_0 m_K^{\rm phys}=1.2544$, there is no need to consider other
values of the quark mass to inter/extrapolate the kaon point.\footnote{The value of the reference scale $r_0$ is set to $r_0=0.5~{\rm fm}$, and we take the ratio $r_0/a$ from~\cite{Necco:2001xg}.}
Again, finite volume effects are
expected to lie within the quoted uncertainties.

Finally, by combining the continuum limit results of 
Eqs.~(\ref{eq:RGI1}-\ref{eq:RGI3}) with the bare Neuberger fermions results
of \ret{tab:overlap_results} we derive non-perturbative estimates 
of the RGI renormalisation factors
\begin{gather}
\left.\frac{\zrgi_1^\pm}{\ZA^2}\right|_{\beta=5.8485} =
\frac{\orgi{\cR}_\pm}{\cR_\pm^{\rm ov}} \, ,~~~~~~~~~
\left.\frac{\zrgi_1^-}{\zrgi_1^+}\right|_{\beta=5.8485} =
\frac{\cR_+^{\rm ov}/\cR_-^{\rm ov}}{\orgi{\cR}_+/\orgi{\cR}_-} \, .
\end{gather}
The results are collected in the last column of \ret{tab:renorm},
together with the corresponding perturbative estimates, which
will be discussed in the next section.

\begin{table}[!t]
\begin{center}
\begin{tabular}{c c c c}
\Hline
\noalign{\vskip0.5ex}
    & bare P.T. & MFI P.T. & non-perturbative\\
\noalign{\vskip0.5ex}
\hline
\noalign{\vskip0.3ex}
$\zrgi_1^{+}/Z_{\rm A}^2$     & 1.242 & 1.193 & 1.15(12)\\
$\zrgi_1^{-}/Z_{\rm A}^2$     & 0.657 & 0.705 & 0.561(61) \\
$\zrgi_1^{-}/\zrgi_1^{+}$ & 0.525 & 0.582 & 0.584(62) \\
\noalign{\vskip0.3ex}
\Hline
\end{tabular}
\caption{Perturbative and non-perturbative estimates for Neuberger fermions RGI
  renormalisation factors at $\beta=5.8485$.}
\label{tab:renorm}
\end{center}
\end{table}

\section{Perturbative estimates of renormalisation factors}

In this section we will determine the RGI renormalisation
factors of interest in perturbation theory. This provides a
handle on the systematics related to their non-perturbative
determination.

The anomalous dimensions $\gamma^\pm$ of the operators $\cQ_1^\pm$ 
are known at two loops for several schemes. For
discretisations based on the Neuberger-Dirac operator, the renormalisation
factors $Z_{\rm s}(g_0,a\mu)$ have been computed for ${\rm s}={\rm
RI/MOM}$ in perturbation theory at one loop in~\cite{Capitani:2000bm}.
The ratios of renormalisation constants we are interested in, computed
with Neuberger fermions and in the RI/MOM scheme, can be written as
%
%
%
\begin{gather}
\label{eq:PTren}
\begin{split}
 \frac{Z^{\pm}_{\rm RI}(g_0,a\mu)}{\ZA^2(g_0)} &=
   1 + (1 \mp 3)\frac{g_0^2}{16\pi^2}\left\{ 2\ln(4\mu{a})
                                 -\frac{1}{3}(B_{\rm S}-B_{\rm V}) \right\}
   +\rmO(g_0^4) \,,\\
 \frac{Z^{-}_{\rm RI}(g_0,a\mu)}{Z^{+}_{\rm RI}(g_0,a\mu)} &=
   1+\frac{g_0^2}{16\pi^2}\left\{ 12\ln(4\mu{a})
                                 -2(B_{\rm S}-B_{\rm V}) \right\}
   +\rmO(g_0^4) \,.   
\end{split}
\end{gather}
It is also possible to perform the expansion using ``mean-field improvement'' (MFI) \cite{Lepage:1992xa}, which aims at improving the convergence
of the perturbative series. At the level of the ratios in Eq.~(\ref{eq:PTren}),
it is easy to check that the implementation of MFI simply amounts to replacing
the bare coupling $g_0^2$ by a ``continuum-like'' coupling $\tilde{g}^2$,
which we set to be $\gbar^2_{\MSbar}$.


The coefficients $B_{\rm S}$ and $B_{\rm V}$ in
Eq.~(\ref{eq:PTren})
are listed in Table~1 of~\cite{Capitani:2000bm}. In order to obtain the corresponding RGI
renormalisation factors, it is enough to multiply the above by the 
suitable perturbative running factors $c_{\rm RI}^\pm(\mu/\Lambda)$.
In our simulations we use $\beta=6/g_0^2=5.8485$.
For $\mu=2\,\GeV$ and $\Lambda=238\,\MeV$~\cite{Capitani:1998mq}, 
the NLO perturbative values for the coefficients $c_{\rm RI}^\pm$ 
are $c_{\rm RI}^{-}(\mu/\Lambda) = 0.6259$ and 
$c_{\rm RI}^{+}(\mu/\Lambda) = 1.2735$. Putting this together
with
Eq.~(\ref{eq:PTren}),
we obtain for the
RGI renormalisation factors the values quoted in the first two
columns of Table~\ref{tab:renorm}. It is worth mentioning that
the differences between perturbative results evaluated in ``bare''
and MFI perturbation theory are relatively
small. This is presumably a consequence of having considered
ratios of operators, in which contributions of the self-energy type cancel,
and is in stark contrast to the situation
encountered in simple quark bilinears, where the deviations
between perturbative and non-perturbative estimates amount to
about 30\% at similar values of the bare coupling~\cite{Hernandez:2001yn,Giusti:2001pk,Wennekers:2005wa}.

This analysis implies, furthermore, that it is unlikely that our
non-perturbative results are affected by large cutoff effects, e.g. those proportional to powers of the quark mass.
\section{Conclusions}

In this work we have laid out a general strategy for the
non-perturbative renormalisation of operators with Neuberger fermions,
via a matching to results obtained with Wilson-type regularisations.
As an application, we have dealt with the overall logarithmic renormalisation
of the operators entering the $\Delta S=1$ effective Hamiltonian
with an active charm quark, for which we have computed RGI 
renormalisation factors in the quenched approximation.
An immediate application of our results appears in the determination 
of the effective couplings governing kaon decays in the low-energy description of the theory~\cite{prl}, in the spirit of~\cite{Giusti:2004an}.

There are a few caveats in this approach:

\begin{itemize}

\item From the technical point of view, we believe that our tmQCD results for the RGI $\orgi{\cQ}^\pm$ constitute a significant advance with respect to previous computations, in that they have been achieved with two Wilson-type regularisations, non-perturbative renormalisation and RG running, at several bare couplings etc. In spite of this, the fact that continuum limit extrapolations are rather long renders the absence of $\Oa$ improvement an important drawback in our effort to obtain stable continuum limit results.
As far as our Neuberger fermions computations are concerned, we point out that, at present, we have results only at one bare coupling. Furthermore, exploring the dependence of renormalisation factors on the choice of reference point would be useful to quantify the impact of ${\rm O}((am)^2)$ cutoff effects.

\item From the conceptual point of view, the specific matching procedure adopted here is based on fixing the matrix element $\langle\pi|\orgi{\cQ}^\pm|K\rangle$ (at $m_\pi=m_K=m_K^{\rm phys}$) to the value predicted with tmQCD Wilson-type fermions. Having used this ``physical'' predictions as renormalisation conditions (for $\orgi{\cQ}^+$ it is the value of the kaon mixing parameter $B_K$) implies that our measurements of $\zrgi_1^+/\ZA^2$ cannot be used for the independent renormalisation of $B_K$ or of $K\to\pi$ matrix elements with Neuberger fermions. On the other hand, our renormalisation
constants are perfectly suitable to renormalise $K\to\pi\pi$ matrix
elements computed in infinite volume and for particle masses in
the physical range, or for ratios of correlation functions computed
in the $\epsilon$-regime of QCD.

\end{itemize}


The ideal approach to the renormalisation problem in hand would involve a working
formulation of the Schr\"odinger functional for Neuberger fermions.
An important recent step in that direction is the proposal of
ref.~\cite{Luscher:2006df}.

\vskip 7truemm
The tmQCD data reported in this work were obtained within an ALPHA Collaboration project; P.D., F.P., C.P. and A.V. wish to thank J. Heitger and S. Sint for discussions.
Our calculations were performed on the APEMille installation of DESY-Zeuthen and on PC clusters at DESY-Hamburg, CILEA and the University of Valencia, as well as on the IBM Regatta at FZ J\"ulich and on the IBM MareNostrum at the Barcelona Supercomputing Center. We thank all these institutions and the University of Milano-Bicocca (in particular C. Destri and F. Rapuano) for their support. P.H. acknowledges partial support by CICYT (grants FPA2004-00996 and FPA2005-01678) and Generalitat Valenciana (GV05-164). F.P. acknowledges the Alexander-von-Humboldt Stiftung for financial support.


\newpage

\end{document}